\documentclass{ws-p8-50x6-00}

\def\jnl#1#2#3#4{{#1} {\bf #2}, #3 (#4)}

\def\npa{{ Nucl. Phys.} A}
\def\plb{{ Phys. Lett.} B}

\newcommand{\xfig}[1]{\begin{center}
\epsfxsize .95\textwidth 
\mbox{\epsffile{#1}}
\end{center}}

\begin{document}

\title{pQCD Structure and Hadronization in Jets and Heavy-ion Collisions}

\author{Thomas A. Trainor}

\address{Nuclear Physics Laboratory 354290, University of Washington  \\ 
 Seattle, WA 98195, USA
\\ 
E-mail: trainor@hausdorf.npl.washington.edu}

\maketitle

\abstracts{Implications of hadronization as a rapid traversal of the QCD phase boundary are explored for correlation structures in jets, N-N collisions and heavy-ion collisions. Hadronization viewed as a partition of the prehadonic system restricts the scale interval over which power-law jet correlation structures survive. Rapid traversal of the QCD phase boundary may result in a lattice-like residual structure in N-N and A-A conifiguration space accessible to momentum-space correlation analysis {\em via} Hubble flow. 
}

\section{Introduction}

The study of pQCD correlation structure in jets at LEP and HERA considers the LPHD hypothesis and the effects of hadronization on expected pQCD gluon bremsstrahlung correlations. Closely-related issues emerge in the treatment of correlations and fluctuations in heavy-ion collisions at the SPS and RHIC and charge-dependent correlations previously observed in the axial phase space of N-N collisions. 

A recent Delphi factorial-moment jet correlation analysis indicates disagreement with the LPHD hypothesis, possibly due to hadronization.  I consider the hadronization process as a rapid traversal of the QCD phase boundary and examine the consequences for correlations and fluctuations. I introduce a generalized treatment of transverse flow effects in nuclear collisions and describe newly observed large-scale momentum correlation structures  which hint at a regular fm-scale space structure in heavy-ion collisions

\section{Jet Correlations}

Jet correlation studies may be applied to gluon bremsstrahlung correlations. The tree-like bremsstrahlung should produce characteristic power-law behavior (intermittency) in correlation measures such as the normalized factorial moments. Analysis results depend strongly on the dimensionality of the analysis space. Projections to lower dimension tend to reduce correlation effects. In a 1D Delphi intermittency analysis\cite{delphi} one observes little evidence for extended bremsstrahlung structure. The result can be interpreted as indicating large-scale jet width fluctuations. 

The hadronization process acts to restrict the scale interval over which the double log approximation (DLA) and local parton-hadron duality hypothesis (LPHD) are valid. The LPHD hypothesis implies point-to-point transfer of correlation structure from parton to hadron system, whereas the Delphi result suggest a sharp scale restriction -- correlation transfer over a restricted scale interval -- implying that hadrons act as partition elements: small-scale glue structure is reabsorbed into finite-size hadrons.  The hadronization process is not a point-coalescence process but instead has finite spatial extent. I now consider hadronization as rapid traversal of the QCD phase boundary.

\section{Rapid Traversal of the QCD Phase Boundary}

\begin{figure}[th]
%\figurebox{20pc}{15pc}{} 
\begin{tabular}{cc}
\begin{minipage}{.57\linewidth}
\xfig{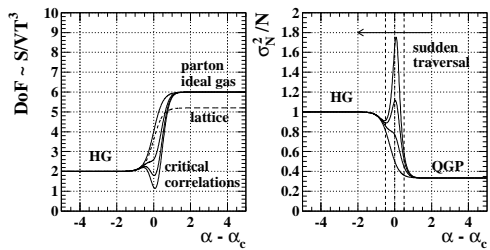}
%\caption{}
\end{minipage}
\begin{minipage}{.37\linewidth}
%\epsfxsize 1\textwidth
%\xfig{totalvar2.eps}  
\caption{Distribution of kinetic degrees of freedom (left) and relative variance (right) on a parameterized thermodynamic state space. Critical correlations formed at the phase boundary may result in reduced DoF density and corresponding increased critical fluctuations.\label{boundary}}
\end{minipage} &
\end{tabular}
%\caption{\label{}}
\end{figure}

In  Fig. \ref{boundary} an idealized phase boundary separates regions with different densities of DoF (left panel). At the boundary large-scale correlations may cause increased critical fluctuations. Depending on the speed of traversal a dynamical system may `remember' more or less about the two boundary aspects: DoF density difference and critical correlations. Fluctuation amplitudes are illustrated in the right panel. We ask whether a system `remembers' a prehadronic correlation state after sudden traversal\cite{raja}.  Increased {\em and} decreased variance depend on traversal rate and {\em observation scale}\cite{tatclt,tatsomething}.  {\em Suppression} of hadronic fluctuations ({\em e.g.,} momentum, flavor, charge, baryon number) implies  an {\em anticorrelated} distribution. Such a distribution must approximate a lattice in configuration space. Rapid traversal of the QCD phase-boundary should be indicated by a {\em local lattice structure} in the distribution of final-state hadrons. 
%%%%%%%%%%%%%%%%%%%%%%%%%%%%%%%
%%%%%%%%%%%%%%%%%%%%%%%%%%%%%%%
Suppressed isovector fluctuations (fluctuations in net pion charge) may reflect a chirally symmetric precursor system, and approach to an {\em isospin antiferromagnet}: two pion charge species form individual  like-sign distributions approximating lattices juxtaposed to provide unlike-sign nearest neighbors.  This picture is supported by the correlation systematics for charged hadrons observed in N-N data.

\section{Phase-space Correlations in N-N and A-A Collisions}

\begin{figure}[th]
\begin{tabular}{cc}
\begin{minipage}{.57\linewidth}
\xfig{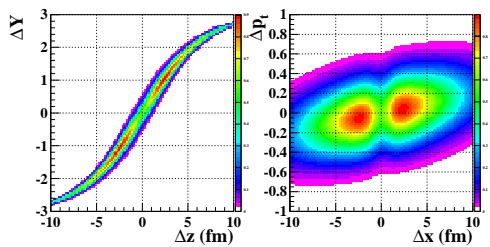}  
%\caption{\label{}}
\end{minipage} &
\begin{minipage}{.37\linewidth}
\epsfysize 1\textwidth
%\pawplot{.ps}  
\caption{Autocorrelation densities for N-N axial phase space (left) and HI transverse phase space (right) illustrating strong (superthermal) configuration-momentum space correlation in the former case and weak (subthermal) correlation in the latter case. \label{g-qy}}
\end{minipage}
\end{tabular}
%\caption{\label{}}
\end{figure}

N-N axial phase space contains very strong ${x}$-$\,{p}$ correlations due to axial Hubble flow (Bjorken expansion) shown in the left panel of Fig. \ref{g-qy}. Correlations in configuration space ($z$) are observable directly in momentum space ($Y$), giving direct access to small-scale configuration-space structure. Correlation structures {\em are} observed\cite{pp} in the two-particle momentum density $\rho_2(y_1,y_2)$ plotted on rapidity difference and azimuth-angle difference. Correlations in N-N collisions indicate substantial anticorrelation of like-sign hadrons and correlation of unlike-sign hadrons at small length scale, consistent with formation of  a 1D lattice structure during hadronization.

A central issue for HI collisions is the correlation structure of  {\em transverse} phase space. Do initial-state strings cross couple\cite{paj}, are color and other measures transversely deconfined? Evidence for small-scale (anti)correlation or lattice structure in HI collisions would be most significant if observed in the transverse phase space. We require  a means to probe directly the local structure of  hadronic space correlations. 
%%%%%%%%%%%%%%%%%%%%%%%%%%%%%%%
%%%%%%%%%%%%%%%%%%%%%%%%%%%%%%%

Fig. \ref{g-qy} contrasts autocorrelation densities for N-N axial phase space and HI transverse phase space at freezeout. Whereas the N-N case shows  a large ${x}$-$ \,{p}$ correlation due to {\em superthermal} Hubble flow the autocorrelation density on heavy-ion transverse phase space shows only a small correlation due to {\em subthermal} Hubble flow. Despite the weak correlation in the heavy-ion case configuration-space correlations may still be observed in momentum space. I now generalize the relationship between collective flow and two-particle momentum correlations.

\section{Two-particle Correlations and Hubble Flow}

To extend the description of transverse flow effects we assume that {pair emission is not random but instead depends locally on partner separation distance}. Guided by FSI treatments\cite{uli1} we factor the two-particle emission density $g(x_1,x_2,p_1,p_2)$ into dependence on mean pair position $x$ and on partner separation $y$. The Wigner density is then
\begin{eqnarray}
S_2(x_1,x_2,p_1,p_2) &\approx& g_+(x,k) \, g_-(y,q)  \\  \nonumber
&\cdot&  f_1(x_1,p_1) \, f_1(x_2,p_2).
\end{eqnarray}
We represent the two-point momentum distribution $P_2({\bf p}_1,{\bf p}_2)$ as a sum of two terms $ A({\bf k,q}) + B({\bf k,q}) $, where $B({\bf k,q}) $ contains the usual HBT Fourier transform of no further interest to us here. $ A({\bf k,q})$ is conventionally assumed to be the product of two single-particle distributions. 

We approximate radial flow locally by analogy with cosmological Hubble flow and assume the form \mbox{ \boldmath $\nabla \beta$} $ = H$\mbox{ \boldmath $\cal I$} from which \mbox{$\Delta$ \boldmath $\beta$}$(x) \approx H \, \Delta {\bf x}$. We assume a Cooper-Frye form for the flow-dependent momentum density. Integrating the Wigner density we obtain the non-interference term of the two-particle momentum distribution as the product of three factors\cite{tathubble}
\begin{eqnarray} \label{threefac}
A({\bf k,q})  &\approx& m_{t1} \, m_{t2} \, e^{-{m_{t1} + m_{t2} \over T}} \\ \nonumber
&\cdot&\!\int\!d^4x g_+(x,k) \, exp\left( {2 H \over T }\, {{\bf k} \cdot {\bf x} } 
\right)   \\ \nonumber
&\cdot&\!\int\!d^4y \,   g_-(y,q) \, cosh\left({{H \over 2 T}  \, {{\bf q} \cdot {\bf y} } } \right),
\end{eqnarray} \nonumber
where $x = (x_1 + x_2)/2$, $y = y_1 - y_2$, are mean and relative four-positions, $k = (p_{1} + p_{2}) / 2$, $q = p_{1} - p_{2}$ are mean and relative four-momenta, $H$ is the local transverse-flow Hubble constant and $T$ is the local temperature.
%%%%%%%%%%%%%%%%%%%%%%%%%%%%%%%
\begin{figure}[th]
\begin{tabular}{cc}
\begin{minipage}{.57\linewidth}
\xfig{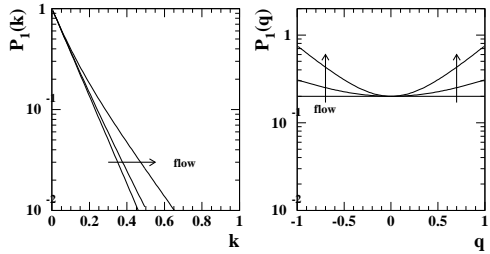}  
%\caption{\label{}}
\end{minipage} &
\begin{minipage}{.37\linewidth}
\epsfysize 1\textwidth
%\pawplot{.ps}  
\caption{Projections of two-particle momentum distribution $P_2({\bf p}_1,{\bf p}_2)$ onto pair mean momentum (left) and pair momentum difference (right) illustrating the effects of transverse Hubble flow on these distributions.\label{hubble6}}
\end{minipage}
\end{tabular}
%\caption{\label{}}
\end{figure}
%%%%%%%%%%%%%%%%%%%%%%%%%%%%%%%

The first two factors represent the Boltzmann distribution on pair mean momentum with the conventional radial-flow-induced blue shift on mean pair momentum $k_t$. The third factor represents a newly-identified quadratic dependence on momentum difference $q$ resulting from transverse Hubble flow. The blue shift and quadratic dependence are represented respectively in the left and right panels of Fig. \ref{hubble6}. The third factor contributes a positive or negative quadratic offset to the two-particle correlator $C_2(k,q)$ on pair-momentum difference if {\em both} Hubble flow and sibling/mixed-pair space-time separation differences are present.

\section{Two-particle  $p_t$ Distributions}

%%%%%%%%%%%%%%%%%%%%%%%%%%%%%%%
\begin{figure}[th]
\begin{tabular}{cc}
\begin{minipage}{.42\linewidth}
\xfig{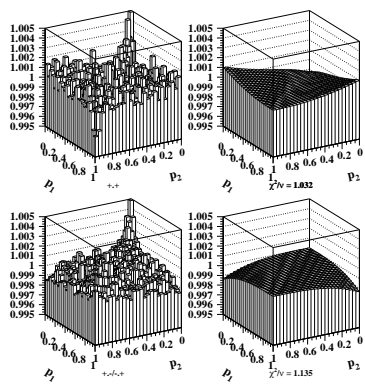}  
%\caption{\label{}}
\end{minipage} &
\begin{minipage}{.52\linewidth}
%\epsfysize 1\textwidth
%\pawplot{.ps}  
\caption{Two-point analysis of $m_t \otimes m_t$ correlations for Pb-Pb collisions at $\sqrt{s_{NN}} = 17~GeV$ (NA49) showing fits (right) to large-scale isospin-dependent correlations for like-sign (top) and unlike-sign (bottom) hadron pairs.\label{ppmfit}}
\end{minipage}
\end{tabular}
%\caption{\label{}}
\end{figure}
%%%%%%%%%%%%%%%%%%%%%%%%%%%%%%%

Two-particle $p_t$ distributions for Pb-Pb collisions at $\sqrt{s_{NN}} = 17~GeV$ (NA49) in Fig. \ref{ppmfit} show significant structure in the pair density ratio. The large-scale quadratic structures (represented by parameterized fits in the right panels) are not related to quantum correlations or tracking effects. Like-sign pairs (top) exhibit an anticorrelated (concave up) distribution on momentum difference, whereas unlike-sign pairs (bottom) exhibit a correlated (concave down) distribution. Referring to Eq. (\ref{threefac}) we see that these results indicate anticorrelation of like-sign pairs and correlation of unlike-sign pairs in transverse configuration space --  formation of a lattice structure -- an {\em isospin antiferromagnet}.
%\section{Pion Lattice Simulations}
We have tested this interpretation with a MC simulation\cite{ronsim} using a transverse lattice to model a boost-invariant system.  Transverse momenta from the measured NA49 $p_t$ spectrum were randomly assigned to particles. A radial flow field also consistent with NA49 measurement was superimposed giving good qualitative agreement with the experimental results. 

%%%%%%%%%%%%%%%%%%%%%%%%%%%%%%%
%\begin{figure}[th]
%\begin{tabular}{cc}
%\begin{minipage}{.57\linewidth}
%\epsfysize .95\textwidth
%\xfig{allsign.eps}  
%\caption{\label{}}
%\end{minipage} &
%\begin{minipage}{.37\linewidth}
%\epsfysize 1\textwidth
%\pawplot{.ps}  
%\caption{Results of a $Q_{inv}$ analysis of lattice-simulation data. These results agree qualitatively with experimental results for central Pb-Pb collisions, suggesting the formation of lattice-like structures in the transverse configuration space of these collisions.\label{allsign}}
%\end{minipage}
%\end{tabular}
%\caption{\label{}}
%\end{figure}
%%%%%%%%%%%%%%%%%%%%%%%%%%%%%%%

\section{Conclusions}

Hadronization modeled as rapid traversal of the QCD phase boundary implies related correlation effects in jets, N-N axial phase space and A-A axial and transverse phase space. Formation of a transverse lattice structure at the $fm$ length scale in high-energy heavy-ion collisions should be possible based on observations of a 1D axial lattice structure in N-N collisions provided there is rapid hadronization from a prehadronic state. A recent two-particle correlation analysis in the transverse phase space of Pb-Pb collisions at the SPS has revealed isospin-dependent correlation structures which suggest that a combination of transverse Hubble flow and $fm$-scale space lattice structure produces observed large-scale momentum correlations. 

Scale-dependent jet correlation studies combined with small-scale probes of the N-N and heavy-ion hadronic freezeout surface  {\em via} Hubble flow and precision analysis techniques provides new access to the small-scale structure of the chemical freezeout surface and the QCD phase boundary. We search for definitive evidence in HI collisions for a deconfined 3D partonic system. Evidence for a pion lattice in {\em transverse} phase space may offer conclusive evidence for a locally-equilibrated prehadronic state in HI collisions.

\end{document}